\documentclass{article}
\usepackage[utf8]{inputenc}
\usepackage{bm}
\usepackage{setspace}
\usepackage{caption}
\usepackage{amssymb}
\usepackage{multirow}  
\usepackage[T1]{fontenc}
\usepackage{lscape}
\usepackage{amsmath}
\usepackage{multirow}
\usepackage{array}
\usepackage{graphicx} 
\usepackage{epstopdf}
\usepackage[shortlabels]{enumitem}
\usepackage{tikz}
\usetikzlibrary{shapes.geometric, arrows,calc}
\usetikzlibrary{arrows,positioning}
\usepackage{natbib}
\usepackage{mathrsfs}
\usepackage{verbatim}
\usepackage{subfig}
\usepackage{geometry}
\usepackage{pdflscape}
\usepackage{hyperref}
\usepackage{amsfonts}

\usepackage{mathrsfs}
\usepackage{booktabs}
\usepackage{pifont}
\usepackage{algorithmicx}
\usepackage{algorithm}
\usepackage{algpseudocode}
\usepackage{changepage}
\usepackage{etoolbox}
\usepackage{appendix}
\usepackage{capt-of}
\usepackage{bbm}
\usepackage{forest}
\usepackage{algpseudocode,algorithm}
\usepackage{authblk}

\title{Structural randomised selection}
\author{Fan Wang$^1$,\ \  Sylvia Richardson$^{1,}$\thanks{Joint Corresponding Author.}  ,\ \  Steven M.\ Hill$^{1,*,\!\!\!}$ \thanks{Present Address: Cancer Research UK Manchester Institute Cancer Biomarker Centre, University of Manchester, Manchester, UK.} \\
$^{1}$MRC Biostatistics Unit, University of Cambridge, Cambridge, UK.
}

\date{}
\usepackage{natbib}
\usepackage{graphicx}

\begin{document}

\maketitle

\begin{abstract}
An important problem in the analysis of high-dimensional omics data is to identify subsets of molecular variables that are associated with a phenotype of interest. This requires addressing the challenges of high dimensionality, strong multicollinearity and model uncertainty. We propose a new ensemble learning approach for improving the performance of sparse penalised regression methods, called STructural RANDomised Selection (STRANDS). The approach, that builds and improves upon the Random Lasso method, consists of two steps. In both steps, we reduce dimensionality by repeated subsampling of variables. We apply a penalised regression method to each subsampled dataset and average the results. In the first step, subsampling is informed by variable correlation structure, and in the second step, by variable importance measures from the first step. STRANDS can be used with any sparse penalised regression approach as the ``base learner''. Using synthetic data and real biological datasets, we demonstrate that STRANDS typically improves upon its base learner, and that taking account of the correlation structure in the first step can help to improve the efficiency with which the model space may be explored.
\end{abstract}

\section{Introduction}

Advancements in high-throughput technologies have resulted in the generation of large collections of high-dimensional omics datasets. For example, the Pan-Cancer Atlas \citep{Hoadley2018} from The Cancer Genome Atlas (TCGA) comprises genomic, transcriptomic, epigenomic and proteomic data across 33 cancer types. It is often of interest to model the relationships between molecular entities measured in such datasets and a phenotype of interest. This is a challenging problem due to, for example, the high dimenionality of the data where there is typically substantially fewer samples than number of variables; high levels of noise can weaken the signals of interest; and complex multicollinearity structure makes it more difficult to distinguish among tightly related variables.

Linear regression models are frequently used to discover relationships between molecular features and a response variable, with variable selection used to discover subsets of features that are particularly relevant for predicting the response. Enforcing sparsity constraints is useful for interpretability and also provides the necessary regularisation to fit models in these high-dimensional settings.
Some popular $l_1$-type regularisation approaches include Lasso \citep{Tibshirani1996}, Elastic Net \citep{Zou2005} and Adaptive Lasso \citep{Zou2006}. Despite their beneficial properties, each method has its drawbacks. Specifically, Lasso and Adaptive Lasso can not select more variables than the sample size, and when multiple variables are highly correlated with each other, they typically select only one or a few of the variables. Elastic Net has a grouping effect property where highly correlated variables tend to have similar coefficient estimates, but this tends to provide inaccurate estimation if highly correlated variables have coefficients of different magnitudes, or even different signs, which is common in biological research. For example, groups of genes may highly coexpress, and expression levels within such a gene block can be highly correlated; however, the effects of these genes on the phenotype may have opposite directions. 


~

\citet{Wang2011} proposed the Random Lasso approach which combines bootstrapping of samples with random subsampling of variables (similar to the random forest method) within a two-step procedure to address the above issues with Lasso, Adaptive Lasso and Elastic Net. 
However, the approach has a number of limitations: it is very computationally intensive, particularly for high-dimensional data; it applies an arbitrary threshold to determine the final set of selected features; and correlation structure between variables is not taken into account when exploring the model space.

We propose a new modelling strategy, STructural RANDomised Selection (STRANDS), that is informed by and improves upon Random Lasso by addressing the above limitations. The method takes account of correlation structure of the data, to explore the model space in a structured way. It has fewer tuning parameters than Random Lasso, and so is computationally more efficient. It uses full sample information (instead of bootstrapping) and its thresholding rule is based on selection probabilities rather than sample size. 

The remainder of the paper is organised as follows. In Section \ref{section:methods} we describe and motivate the proposed method. In Section \ref{subsection:low_dimension} and \ref{subsectio:high_dimensional} we show results comparing STRANDS with Random Lasso and other popular Lasso variants using both low-dimensional and high-dimensional datasets. In Section \ref{section:further_analysis_strands} we investigate the effects of clustering step in STRANDS. In Section \ref{sec:real_data} we show results using real data. We conclude with a discussion in Section~\ref{section:discussion}.

\section{Methods} \label{section:methods}
Assume the data consists of $n$ samples and $p$ variables. $\textbf{Y}\in \mathbb{R}^n$ is a $n\times 1$ response vector, $\bf{X}$ is a $n\times p$ design matrix of $p$ variables $(\boldsymbol{x_1} ,\dots, \boldsymbol{x_p}), \boldsymbol{x_j}\in \mathbb{R}^n$ for $j=1\dots p$. The Gaussian linear regression model takes the form

\begin{equation} \label{eq:linear_model}
\textbf{Y}=\textbf{X}\boldsymbol{\beta}+\boldsymbol{\epsilon},
\end{equation}

\noindent where $\boldsymbol{\beta}=(\beta_1 ,\dots, \beta_p)^T$ is a $p\times 1$ vector of coefficients, and $\boldsymbol{\epsilon}\sim N(\textbf{0},\sigma^2 \textbf{I})$ is a $n \times 1$ vector of error terms, $\textbf{I}$ being the identity matrix. We assume that each column of design matrix $\textbf{X}$ has been standardised to have mean zero and variance one, and $\textbf{Y}$ is mean-centred. This results in the intercept term being zero and it is therefore omitted from model (\ref{eq:linear_model}). We further define $s_0$ as the number of non-zero coefficients.

Penalised regression methods augment the loss function with a penalty term that encodes a structural assumption such as sparsity. Specifically, in penalised likelihood regression, one minimises the objective function 
\begin{equation} \label{eq:penalised_linear_model}
M(\boldsymbol{\beta})=||\textbf{Y}-\boldsymbol{X\beta}||_2^2+P_\lambda(\boldsymbol{\beta}),
\end{equation} where $P_\lambda(.)$ is a penalty function that penalises complex models, and $\lambda>0$ controls the degree of penalisation. $P_\lambda(\boldsymbol{\beta})$ shrinks some coefficient estimates towards zero, and when $P_\lambda(\boldsymbol{\beta})$  satisfies certain properties, some of the coefficients are shrunk to exactly zero, leading to a sparse solution.

\subsection{Lasso}\label{subsec:lasso}
The Lasso (least absolute shrinkage and selection operator) estimator \citep{Tibshirani1996} takes the form given in (\ref{eq:penalised_linear_model}) with an $l_1$-norm penalty:
$P_{\lambda}(\beta_j)=\lambda|\beta_j|$.
This shrinks coefficients towards zero, with some set to exactly zero, and $\lambda$ controls the amount of shrinkage and degree of sparsity. Lasso can only select up to $n$ variables, and does not have group selection property, such that it tends to only select a few variables from a group of highly correlated variables. The level of regularization $\lambda$ is usually determined to minimise cross-validation error in practice. 

\subsection{Elastic Net}
Elastic Net \citep{Zou2005} minimises (\ref{eq:penalised_linear_model}) with a convex combination of $l_1$ and $l_2$ penalty $P_{\lambda}(\beta_j)=\lambda_1|\beta_j|+\lambda_2\beta_j^2=\lambda \left(\alpha |\beta_j|+\frac{1-\alpha}{2}\beta_j^2\right)$, where $\lambda_1, \lambda_2>0$ and $\alpha \in [0,1]$ controls the balance of the two. The $l_1$ penalty guarantees automatic variable selection and continuous shrinkage, while the $l_2$ penalty helps select correlated variables simultaneously thanks to the group selection property, and stabilises the solution path. 

\subsection{Adaptive Lasso}           
Adaptive Lasso \citep{Zou2006} aims to adaptively impose penalties on different coefficients, such that large coefficients are less penalised than small ones. It uses the following penalty in (\ref{eq:penalised_linear_model})
\begin{equation}
P_\lambda(\beta_j)=\lambda w_j |\beta_j|,
\end{equation} where $w_j>0$ is a known weight for variable $j$. When $w_j$ has the form $\frac{1}{|{\hat{\beta_j}}|^{\tau}}$ for some $\tau>0$ and $\hat{\beta_j}$ is a root-n consistent estimate of $\beta_j$, adaptive Lasso also enjoys the oracle property \citep{Zou2006}.

\subsection{Random Lasso} \label{subsection:random_lasso}
Random Lasso \citep{Wang2011} is a variable selection method motivated by random forest. It consists of two steps and can be summarised as follows: \\
\\
\noindent Step 1. Calculate importance measure \\

1a. For $b_1=1,\dots B$, do: 
\begin{enumerate}[label=(\roman*)]
\item  Draw a bootstrap sample by sampling with replacement from full data 
\item Randomly select $q_1$ variables from the bootstrapped samples and apply Lasso to these variables and obtain coefficient estimates $\tilde{\beta}_j^{(b_1)}, j=1\dots p$. Coefficients for those not among the $q_1$ variables are set to be zero
\end{enumerate}

1b. The importance measure of $x_j$ is calculated as $I_j=|\frac{\sum_{b_1=1}^B {\tilde{\beta_j}^{(b_1)}}}{B}|$, $j=1\dots p$\\
\\
\noindent Step 2. Select variables \\

2a. For $b_2=1,\dots B$, do: 

\begin{enumerate}[label=(\roman*)]
\item Draw a bootstrap sample by sampling with replacement from full data 
\item Randomly select $q_2$ variables from the bootstrapped samples, with selection probability of $x_j$ proportional to its importance measure $I_j$. Apply Lasso to these variables to obtain coefficient estimates $\hat{\beta_j}^{(b_2)}, j=1\dots p$. Coefficients for those not among the $q_2$ variables are set to be zero\\
\end{enumerate}

2b. The final estimator is $\hat{\beta_j}=\frac{\sum_{b_2=1}^B {\hat{\beta_j}^{(b_2)}}}{B}$. \citet{Wang2011} suggest to set the threshold on magnitudes of coefficient estimates to be $1/n$, such that those variables with $|\hat{\beta_j}|<1/n$ are removed from the model. \\

Sampling of variables in Step 1a and 2a breaks down the correlation structure, such that highly correlated variables are selected in different candidate models. In penalised linear regression choosing the right amount of penalisation is notoriously difficult, especially for high-dimensional data, and there may not be a single penalty level that can perfectly recover the sparsity pattern \citep{Meinshausen2010}. Random Lasso applies Lasso on different candidate models, with different penalty levels; the results are summarised across candidate models, with more robust performance. The number of selected variables is no longer limited by $n$. $q_1$ and $q_2$ are tuning parameters, and \citet{Wang2011} suggest using cross-validation to find optimal $q_1$ and $q_2$, by repeated application of Random Lasso. 

Random Lasso manages to solve some issues of Lasso, namely, it can simultaneously select highly correlated variables even if they have coefficients of different signs, and the number of selected variables is not limited by sample size. In spite of the nice properties, it has several potential drawbacks: 
\begin{enumerate}
\item since the search for optimal combination of $q_1$ and $q_2$ is achieved by grid search, the method is rather time consuming, particularly for high-dimensional data; 
\item fixing $q_1$ may not be the best way to fully explore candidate models in Step 1; 
\item the threshold $1/n$ on coefficient estimates is somewhat arbitrary since magnitudes of coefficients do not depend on sample size; 
\item each bootstrapped sample loses partial information of the original data. 
\item the approach does not take account of correlation structure among variables; it is known that understanding the correlation structure before regression could help better distinguish relevant variables from correlated irrelevant ones (see e.g. \cite{Bulmann2013}).
\end{enumerate}

\subsection{STructural RANDomised Selection (STRANDS)}
We propose a new approach, STructural RANDomised Selection (STRANDS), that is similar in spirit to Random Lasso, but aims to address its aforementioned limitations to improve performance.

\begin{figure}[!h]
\centering
\includegraphics[width=6.2in,height=6.2in]{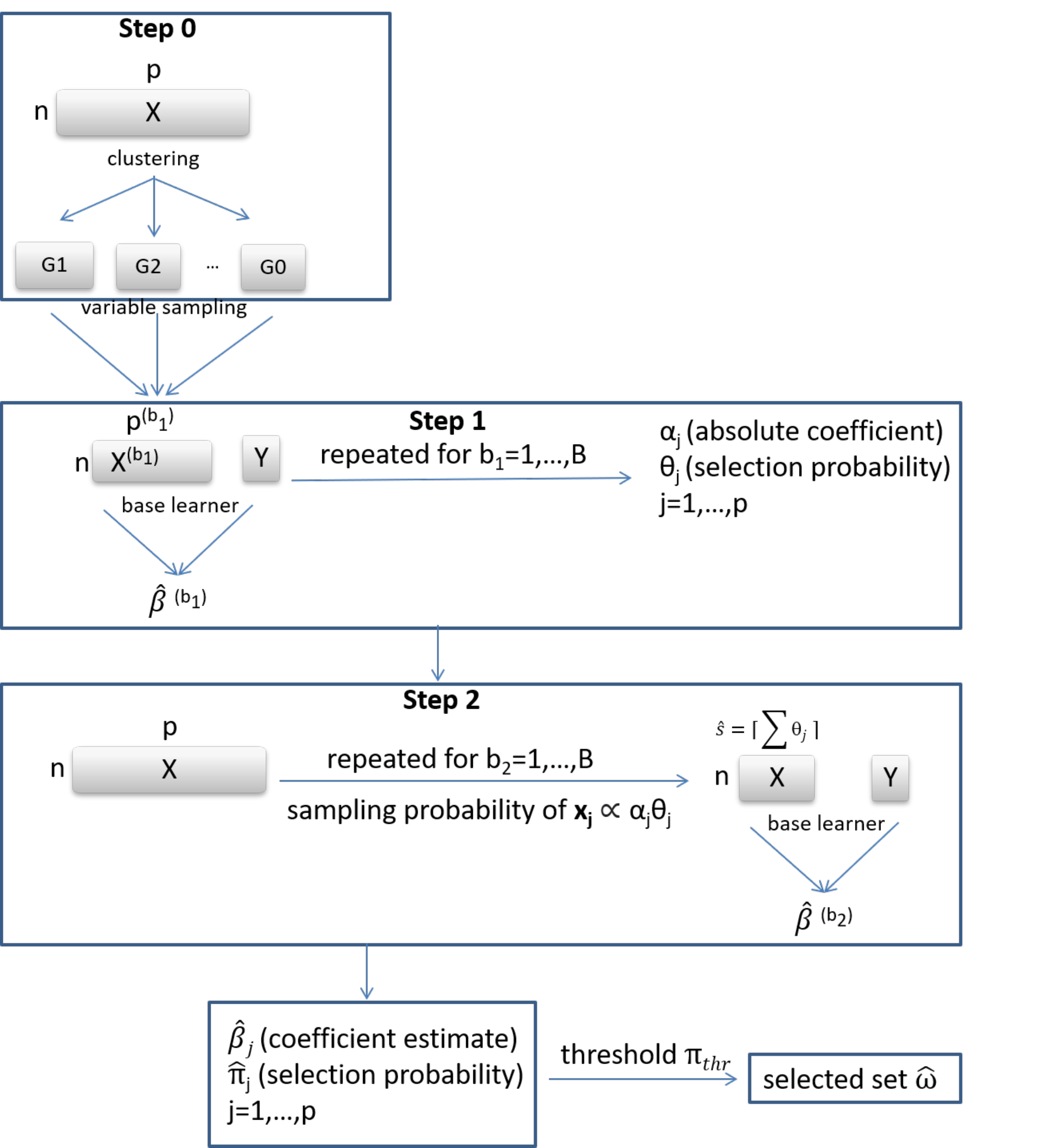}
\caption{An illustration of STRANDS}
\label{fig:strands}
\end{figure}

STRANDS consists of three steps. It first captures the correlation structure of data. Then it randomly selects variables informed by the correlation structure, before performing regression on those variables. With the information from model exploration, the final step aims to randomly remove irrelevant variables while retaining the relevant ones, before performing regression again. The results are averaged across candidate models and the final thresholding rule is based on selection probabilities. STRANDS can be combined with any sparse regression method. We refer to this method as the `` base learner ''. Each step is explained in more detail below and illustrated in Figure \ref{fig:strands}.

\begin{algorithm}
\caption{Correlation clustering algorithm}\label{algorithm:clustering}
\begin{algorithmic} 

\State Input: $n$ by $p$ design matrix $\bf{X}$, response vector $\bf{Y}$, sparse regression base learner algorithm $\mathcal{A}$ and correlation threshold $\rho_0>0$ 
\State
\State Output: An independent group of variables $G_0$ and $K$ correlated groups of variables $G_1 \dots G_K$
\State
\State Initialise $k=1$; correlated variables set $G=\emptyset$; remaining variables set $R=\{\bf{x_1} \dots \bf{x_p} \}$
\State
\State 
\State Run base learner algorithm $\mathcal{A}$ on the whole data 
\State Let the set of selected variables be $\textbf{S}$

\For{$\boldsymbol{x} \in \textbf{S}$}
\State  \If{$\boldsymbol{x} \in G$}
                \State next
                \EndIf 
\State        $G_k=\{\boldsymbol{x}\}$; $\rho_M=1$ 
        \While{$\rho_M \geq \rho_0$}
\State Find the variable $\boldsymbol{x_r} \in R \setminus G_k$ that has the highest median absolute 
\State correlation with elements of $G_k$
\State Set $\rho_M$ to be the highest median absolute correlation
       \If{$\rho_M\geq \rho_0$} 
\State $G_k=G_k \cup \{\boldsymbol{x_r}\}$ 
       \EndIf
\EndWhile  
 
 \If{$|G_k| \geq 2$}
\State $k=k+1$
\State $R=R \setminus G_k$
 \EndIf              
  \EndFor \\
$G_0=R$, $K=k-1$
\end{algorithmic}
\end{algorithm}

\begin{algorithm}
\caption{STructural RANDomised Selection (STRANDS)}
\begin{algorithmic}

\State Input: $n$ by $p$ design matrix $\bf{X}$, response vector $\bf{Y}$, sparse regression base learner algorithm $\mathcal{A}$, correlation threshold $\rho_0>0$, number of iterations B and threshold probability $\pi_{thr}$
\State 
\State Output: coefficient estimate ($\hat{\beta}_j$) and selection probability ($\hat{\pi}_j$) of each variable, and a set of selected variables $\hat{\omega}$
\State
\State Step 0: Apply correlation clustering (Algorithm \ref{algorithm:clustering}), resulting in one independent group ($\boldsymbol{G_0}$) and $K$ groups of correlated variables ($\boldsymbol{G_1}$ $\dots$ $\boldsymbol{G_K}$)
\State
\State Step 1: Calculate importance measure 

\State 1a
\For{$b_1=1,\dots, B$}
\State (i)
       \For {$k=0,\dots ,K$}
\State Sample without replacement from $\boldsymbol{G_k}$ to obtain $\boldsymbol{G_k'}$ $\subseteq$ $\boldsymbol{G_k}$, where $|\boldsymbol{G_k'}|$ 
\State is uniformly drawn from $\{0,1,\dots ,|\boldsymbol{G_k}| \}$
       \EndFor
       \State  $\boldsymbol{S^{(b_1)}}=\bigcup\limits_{k=0}^{K} \boldsymbol{G'_{k}}$. 

\State (ii) Apply base learner algorithm $\mathcal{A}$ to $(\boldsymbol{S^{(b_1)}},\textbf{Y})$, tuned on the whole range 
\State of penalty levels (as $\textit{per}$ default choice of $\mathcal{A}$).  
\State For $\boldsymbol{x_j} \in \boldsymbol{S^{(b_1)}}$, denote the resulting coefficient estimate by $\hat{\beta}_j^{(b_1)}$. 
\State For $\boldsymbol{x_j} \not\in \boldsymbol{S^{(b_1)}}$, set $\hat{\beta}_j^{(b_1)}=0$ 
   \EndFor

\State 1b
\For {$j=1,\dots ,p$}
\State $m_j=\sum\limits_{b_1=1}^B {\mathbbm{1}\{\boldsymbol{x_j} \in \boldsymbol{S^{(b_1)}} \}}$
\State $\alpha_j=\frac{\sum\limits_{b_1=1}^B{|\hat{\beta}_j^{(b_1)}|}}{m_j}$ 
\State $\theta_j=\frac{\sum\limits_{b_1=1}^B{\mathbbm{1}(\hat{\beta_j}^{(b_1)} \neq 0)}}{m_j}$
\EndFor 

\algstore{myalg}
\end{algorithmic}
\end{algorithm}

\begin{algorithm}                     
\begin{algorithmic}                   
\algrestore{myalg}

\State Step 2: Select variables
\State 2a
\For {$b_2=1,\dots, B$}
\State (i) Randomly select $\tilde{s}=\lceil \sum\limits_{j=1}^p{\theta_j} \rceil$ ($\lceil \cdot \rceil$ denotes the ceiling function) of the $p$ 
\State variables, with selection probability of $\boldsymbol{x_j}$ proportional to $\alpha_j \theta_j$, $j=1, \dots ,p$
\State Denote the set of selected variables by $\boldsymbol{S^{(b_2)}}$

\State (ii) Apply base learner algorithm $\mathcal{A}$ to $(\boldsymbol{S^{(b_2)}},\textbf{Y})$, tuned on the set of optimal 
\State penalty levels obtained from Step 0 and 1. 
\State For $\boldsymbol{x_j} \in \boldsymbol{S^{(b_2)}}$, denote the resulting coefficient estimate by $\hat{\beta}_j^{(b_2)}$. 
\State For $\boldsymbol{x_j} \not\in \boldsymbol{S^{(b_2)}}$, set $\hat{\beta}_j^{(b_2)}=0$   
\EndFor 

\State 2b
\For {$j=1,\dots ,p$}
\State $\hat{\beta_j}=\frac{\sum\limits_{b_2=1}^B {\hat{\beta}_j^{(b_2)}}}{B}$
\State $\hat{\pi}_j=\frac{\sum\limits_{b_2=1}^B {I(\hat{\beta}_j^{(b_2)} \neq 0)}}{B}$
\EndFor 
\State Let $\hat{s}_0=\sum\limits_{j} {I(\hat{\pi}_j\geq \pi_{thr})}$
\State Select variables with the top $\hat{s}_0$ largest coefficient estimates $\hat{\beta}_j$ or the top $\hat{s}_0$ largest selection probabilities $\hat{\pi}_j$. Denote the set of selected variables by $\hat{\omega}$
\end{algorithmic}
\end{algorithm}

In Step 0, correlation structure is determined. We first apply the sparse regression base learner to the complete data. Then for each of the selected variables, we form a correlated group by iteratively adding in the variable that is most correlated with existing members of the group. Any variables that do not belong to any of these correlated groups form an `` independent group ''. 
As a result, the $p$ variables are divided into $K$ groups of correlated variables ($G_1, G_2 \dots G_K$), and an independent group $G_0$ (see Figure \ref{fig:strands}). These groups are used in the next step to inform sampling of variables. 

In Step 1, variables are sampled within each group independently. This helps better explore the model space, as shown in Section \ref{subsubsection:structural_subsampling}. After repeatedly applying the sparse base learner to sampled variables, an average coefficient measure $\alpha_j$ and a selection probability $\theta_j$ are obtained for each variable (see Figure \ref{fig:strands}). They are two importance measures quantifying the association between a variable and the response. 

In Step 2, sampling of variables is informed by importance measures from Step 1. In particular, the selection probability of $\boldsymbol{x_j}$ is proportional to $\alpha_j \theta_j$. The sparse base learner is then applied to the sampled variables. Variable sampling and sparse regression are repeated multiple times before calculating average coefficient estimate $\hat{\beta}_j$ and selection probability $\hat{\pi}_j$. Variables with large selection probabilities or coefficient estimates are declared as relevant (see Figure \ref{fig:strands}). Similar as Random Lasso, since correlated variables are separated and can be selected from different candidate models, STRANDS also has group selection property. 

In Step 1 and 2, when sampling variables, instead of tuning the number of sampled variables as in Random Lasso, they are chosen automatically in STRANDS. In Step 1 a random number of variables are drawn from each of the $K+1$ groups that result from the clustering algorithm; In Step 2 the number of selected variables is the sum of selection probabilities from Step 1, which is an estimation of the underlying sparsity. The automatic choices of parameters make STRANDS much more computationally efficient than Random Lasso. Unlike Random Lasso, STRANDS uses full sample information rather than bootstrapped samples. We propose a thresholding rule not depending on sample size $n$ (Random Lasso uses a threshold $1/n$), but on selection probabilities, e.g., variables selected in more than a fraction of sub-models in Step 2 are declared as important; we also keep variables with largest coefficient estimates. Since the selection probabilities of signals tend to be pushed towards 1, while the non-signals towards 0, a threshold of $\pi_{thr}=0.5$ is reasonable, and it seems to work well in practice. In general we recommend $B$ to be a fairly large value (e.g. $B\geq 200$) for effective exploration of model space and stable results. 

Note that in Step 1, we use the grid of tuning parameters of each sub-dataset independently, to find the best penalty level for each sub-model (by e.g. cross-validation). However, in Step 2, using a grid that ignores the existence of previous steps may not be appropriate. Specifically, \cite{Zhu2014} prove that, after screening, one should apply the penalty level optimal for the full data on the survived variables, in order to achieve unbiased estimation. In other words, if we treat the survived variables after screening as if they are given in the first place, the survived variables tend to prefer smaller level of penalty, such that bias can occur. Motivated by this argument, the grid of tuning parameters for sub-models in Step 2 is chosen to be the set of optimal penalty levels from Step 0 and Step 1. The purpose is to mitigate the screening bias, but as we will see in the results, this bias does not altogether disappear. 

\section{Results} \label{section:results}
\subsection{Low-to-moderate-dimensional settings} \label{subsection:low_dimension}
In this section we perform several simulation studies to demonstrate the proposed method and compare to popular variants of Lasso. Data are generated using Model (\ref{eq:linear_model}):  
$\textbf{Y}=\textbf{X}\boldsymbol{\beta}+\boldsymbol{\epsilon}$. 
For all simulation settings, each variable follows a standard normal distribution, but the correlation structure may vary. $\epsilon_i \sim N(0,\sigma^2)$ for $i=1,\dots ,n$. Variables are standardised to have mean zero and variance one, and responses are centred. \\

Example \ref{table:lasso1} and Example \ref{table:lasso2} were used in the Lasso paper \citep{Tibshirani1996}, where correlation structures follow the Toeplitz design. Example \ref{table:extreme_correlation} was used in the Random Lasso paper \citep{Wang2011}, where coefficients of highly correlated variables have different signs. We have $p>n$ in Examples \ref{table:independence} and \ref{table:block}, one with orthogonal variables and the other with pairwise correlation structure. The details of the five examples are as follows, where signal to noise ratio (SNR) is defined as $\mathrm{SNR} = \sqrt{\boldsymbol{\beta}^T \textbf{X}^\text{T}\textbf{X}\boldsymbol{\beta}/(n \sigma^2)}$. \\
\\
Example \ref{table:lasso1}. There are $p=8$ variables, and the pairwise correlation between variables $x_i$ and $x_j$ is $0.5^{|i-j|}$, $i,j=1\dots 8$. The true coefficients are $\boldsymbol{\beta}=(3,1.5,0,0,2,0,0,0)$, $\sigma=3$, SNR $\approx$ 1.5. \\
\\
Example \ref{table:lasso2}. The model is the same as in Example \ref{table:lasso1}, except $\beta_j=0.85$ for $j=1\dots 8$. i.e., a non-sparse model. SNR $\approx$ 1.25. \\
\\
Example \ref{table:extreme_correlation}. There are $p=40$ variables. The first 10 variables are relevant variables, and the correlation between each pair of them is 0.9. The rest 30 variables are independent from each other and independent from the 10 relevant variables. Relevant variables have coefficients 3, 3, 3, 3, 3, -2, -2, -2, -2, -2. $\sigma=3$. SNR $\approx$ 1.78\\
\\
Example \ref{table:independence}. There are $p=300$ variables. All variables are i.i.d. drawn from $N(\bf{0,I})$. 10 relevant variables are randomly allocated among $p$ variables, which have coefficients 3, 3, 3, 3, 3, 4, 4, 4, 4, 4. $\sigma=3$, SNR $\approx$ 3.65\\
\\
Example \ref{table:block}. Same as in Example \ref{table:independence}, but the first 100 variables are 10 blocks of 10 pairwise correlated variables with correlation $\rho=0.7$, with one relevant variable in each block; the ten blocks are independent from each other. The rest 200 variables are independent from each other, and independent from the 10 blocks of correlated variables, with no signals. SNR$\approx$ 3.65.\\

To assess selection performance, we use the number of true positives (TP), number of false positives (FP), $\text{PPV}=\frac{\text{TP}}{\text{TP+FP}}$.  To assess prediction performance, we use mean squared error, defined as MSE=$(\boldsymbol{\hat{\beta}-\beta})^T \bf{V} (\boldsymbol{\hat{\beta}-\beta})$, where $\bf{V}$ is the population covariance matrix of $\bf{X}$. For STRANDS and RLasso, we use the final selected model after thresholding to assess their selection and prediction performance. 

We compare Lasso, Elastic net (ENet, $\alpha=0.5$),  with Random Lasso (RLasso) and STRANDS-Lasso (STRD-Lasso). We use $B=300$, 
$\rho_0=0.5$,  $\pi_{thr}=0.5$ for STRANDS, and for Random Lasso, we use $B=300$ and search a grid of 0, $0.2p$, $0.4p$, $0.6p$, $0.8p$ and $p$ to find the optimal $q_1$ and $q_2$. For all methods penalty level $\lambda$ is chosen by fivefold cross-validation. 
Results are shown in Tables \ref{table:lasso1} $\sim$ \ref{table:block} and are averages across 100 replicates, with bootstrap standard errors in parentheses. We highlight the method with best performance in bold in each row. We separate the comparison of Adaptive Lasso (AdaLasso, Lasso estimates as initial weights). Hereafter when referring to STRANDS, we mean both STRANDS-Lasso (STRD-Lasso) and STRANDS-AdaLasso (STRD-AdaLasso). The results are shown in Tables \ref{table:lasso1} $\sim$ \ref{table:results_low_ada2}. 

\begin{table}[!htp]
\centering
 \begin{adjustwidth}{-0cm}{}
\begin{tabular}{|c|c|c|c|c|c|c|}\hline
$n$ & & Lasso & ENet & RLasso & STRD-Lasso \\\cline{1-6}
 \multirow{4}{*}{$20$}
    & \#FP & 2.28(0.15) & 2.58 (0.16)  & 2.97 (0.12) & \textbf{1.95 (0.12)} \\\cline{2-6}
    & \#TP & 2.73 (0.05) & 2.8 (0.05)  & \textbf{2.87 (0.04)} & 2.77 (0.05)\\\cline{2-6}
    & PPV & 0.6 (0.02) & 0.57 (0.018) & 0.51 (0.014) & \textbf{ 0.62 (0.019) } \\\cline{2-6}
    & MSE &  4.45 (0.35) & 4.40 (0.35)  & 4.16 (0.23) & \textbf{3.64 (0.25)} \\\hline
\multirow{4}{*}{$50$}         
    & \#FP & 2.17 (0.16) & 2.72 (0.14)  & 3.05 (0.13) & \textbf{2.02 (0.13)} \\\cline{2-6}
    & \#TP & 3 (0) & 3 (0)  &3 (0) & 3 (0) \\\cline{2-6}
    & PPV & \textbf{0.64 (0.019)} & 0.56 (0.015)  & 0.52 (0.014) & \textbf{0.64 (0.018)}\\\cline{2-6}
    & MSE & 1.37 (0.09) & 1.40 (0.09)  & 1.32 (0.09) & \textbf{1.18 (0.08)} \\\hline
\end{tabular}
\end{adjustwidth}
\caption {Example \ref{table:lasso1} results: $p=8, s_0=3, \sigma=3, \rho(x_i,x_j)=0.5^{|i-j|}, \boldsymbol{\beta}=(3,1.5,0,0,2,0,0,0)$.}
\label{table:lasso1}
\end{table}

\begin{table}[!htp]
\centering
 \begin{adjustwidth}{-0cm}{}
\begin{tabular}{|c|c|c|c|c|c|}\hline
$n$ & & Lasso & ENet  & RLasso & STRD-Lasso \\\cline{1-6}
 \multirow{2}{*}{$20$}
    & \#TP  & 5.39 (0.18) & 5.84 (0.16)  & \textbf{5.85 (0.13)} & 5.43 (0.14) \\\cline{2-6}
    & MSE  &  4.95 (0.32) & 4.55 (0.31)  & 4.80 (0.19) & \textbf{4.17 (0.23)} \\\hline
\multirow{2}{*}{$50$}         
    & \#TP & 7.44 (0.07) & \textbf{7.54 (0.06)}  & 7.39 (0.07) & 7.44 (0.07) \\\cline{2-6}
    & MSE  & 1.41 (0.07) & \textbf{1.32 (0.06)} & 2.13 (0.09) & 1.38 (0.06) \\\hline
\end{tabular}
\end{adjustwidth}
\caption{Example \ref{table:lasso2} results: Same as Table \ref{table:lasso1}, $\beta_j=0.85$ $\forall j$}
\label{table:lasso2}
\end{table}

\begin{table}[!htp]
\centering
 \begin{adjustwidth}{-0cm}{}
\begin{tabular}{|c|c|c|c|c|c|}\hline
$n$ & & Lasso & ENet  & RLasso & STRD-Lasso \\\cline{1-6}
 \multirow{4}{*}{$50$}
    & \#FP & 4.22 (0.49) & 5.42 (0.54)  & 12.91 (0.36) & \textbf{3.9 (0.27)} \\\cline{2-6}
    & \#TP & 3.24 (0.11) & 4.87 (0.11)  & 6.6 (0.17) & \textbf{6.82 (0.19)} \\\cline{2-6}
    & PPV & 0.59 (0.028) & 0.57 (0.023)  & 0.35 (0.008) & \textbf{0.66 (0.019)} \\\cline{2-6}
    & MSE  & 6.32 (0.19) & 6.36 (0.25)  & 6.16 (0.16) & \textbf{5.22 (0.20)} \\\hline
\multirow{4}{*}{$100$}         
    & \#FP & 10.71 (1.08) & 10.68 (1.01)  & 16.23 (0.33) & \textbf{5.34 (0.41)}  \\\cline{2-6}
    & \#TP  & 5.96 (0.27) & 6.4 (0.23)  & 9.32 (0.08) & \textbf{9.61 (0.08)}  \\\cline{2-6}
    & PPV  & 0.53 (0.026) & 0.53 (0.026)  & 0.37 (0.004) & \textbf{0.69 (0.017)}  \\\cline{2-6}
    & MSE  & 4.92 (0.11) & 4.89 (0.097) & 3.18 (0.12) & \textbf{2.21 (0.10)}  \\\hline
\end{tabular}
\end{adjustwidth}
\caption{Example \ref{table:extreme_correlation} results: $p=40, s_0=10, \sigma=3$. One block of 10 pairwise correlated variables (correlation $\rho=0.9$), all of which are signals, with coefficients 3, 3, 3, 3, 3, -2, -2, -2, -2, -2. The other 30 variables are independent from each other, and independent from the 10 correlated variables, all of which are noise variables.}
\label{table:extreme_correlation}
\end{table}

\begin{table}[!htp]
\centering
 \begin{adjustwidth}{-0cm}{}
\begin{tabular}{|c|c|c|c|c|c|}\hline
$n$ & & Lasso & ENet  & RLasso & STRD-Lasso \\\cline{1-6}
 \multirow{4}{*}{$50$}
    & \#FP & 18.88 (1.28) & 25.95 (1.43)  & 19.28 (0.67) & \textbf{15.08 (0.6)}  \\\cline{2-6}
    & \#TP & 6.98 (0.27) & 7.54 (0.25)  & 7.62 (0.15) & \textbf{7.64 (0.18)} \\\cline{2-6}
    & PPV & 0.34 (0.019) & 0.28 (0.017)  & 0.3 (0.010) & \textbf{0.36 (0.012)} \\\cline{2-6}
    & MSE & 72.39 (3.25) & 74.96 (2.67)  & 67.53 (2.43) & \textbf{61.91 (3.01)} \\\hline
\multirow{4}{*}{$100$}         
    & \#FP & 38.92 (1.4) & 50.79 (1.55) & 47.62 (0.72) & \textbf{23.71 (0.57)}\\\cline{2-6}
    & \#TP & 10 (0) & 10 (0)  & 10 (0) & 10 (0) \\\cline{2-6}
    & PPV  & 0.22 (0.0062) & 0.18 (0.0046) & 0.18 (0.0023) & \textbf{0.52 (0.011)}\\\cline{2-6}
    & MSE & 8.11 (0.33) & 11.32 (0.43) & \textbf{3.65 (0.17)} & 7.39 (0.31) \\\hline
\end{tabular}
\end{adjustwidth}
\caption{Example \ref{table:independence} results: $p=300, s_0=10, \sigma=3$. Variables are i.i.d. drawn from $N\textbf{(0, I)}$. 10 signals are randomly allocated among $p$ variables, with coefficients 3, 3, 3, 3, 3, 4, 4, 4, 4, 4}
\label{table:independence}
\end{table}

\begin{table}[!htp]
\centering
 \begin{adjustwidth}{-0cm}{}
\begin{tabular}{|c|c|c|c|c|c|} \hline
 $n$ & & Lasso & ENet  & RLasso & STRD-Lasso\\\cline{1-6}
 \multirow{4}{*}{50}
    & \#FP & 29.42 (1.17) & 38.6 (1.01) & 26.8 (0.73) & \textbf{23.49 (0.55)}  \\\cline{2-6}
    & \#TP & 6.6 (0.21) & \textbf{7.33 (0.17)}  & 6.45 (0.17) & 7.32 (0.17) \\\cline{2-6}
    & PPV & 0.19 (0.0066) & 0.17 (0.0046)  & 0.2 (0.0067) & \textbf{0.24 (0.0063)} \\\cline{2-6}
    & MSE & 60.32 (2.20) & 58.90 (1.74)  & 64.86 (1.95) & \textbf{51.18 (1.72)} \\\hline
\multirow{4}{*}{100}         
    & \#FP & 41.5 (1.41) & 56.94 (1.42) & 43.62 (0.99) & \textbf{35.13 (0.61)} \\\cline{2-6}
    & \#TP & 9.99 (0.01) & 9.98 (0.01) & 9.98 (0.01) & 9.98 (0.01) \\\cline{2-6}
    & PPV  & 0.21 (0.005) & 0.16 (0.003)  & 0.19 (0.004) & \textbf{0.23 (0.003)} \\\cline{2-6}
    & MSE & 9.67 (0.44) & 13.49 (0.54)  & \textbf{5.26 (0.44)} & 9.94 (0.42) \\\hline
\end{tabular}
\end{adjustwidth}
\caption{Example \ref{table:block} results: $ p=300, s_0=10, \sigma=3$, 10 blocks of 10 pairwise correlated variables (correlation $\rho=0.7$), each having one signal. The other 200 variables are independent from each other, and independent from the 10 blocks of correlated variables, all of which are noise variables. Signals have coefficients 3, 3, 3, 3, 3, 4, 4, 4, 4, 4.}
\label{table:block}
\end{table}

\begin{table}[!htp]
\footnotesize
\begin{tabular*}{\linewidth}{ @{\extracolsep{\fill}} ll *{13}c @{}}
\toprule
$n$ &  & \multicolumn{2}{c}{Example \ref{table:lasso1}} &
\multicolumn{2}{c}{Example \ref{table:lasso2}} \\
\midrule \midrule \addlinespace \\
 & & AdaLasso  & STRD-AdaLasso  & AdaLasso  & STRD-AdaLasso \\
\addlinespace \\
\multirow{3}{*}{20} & \#FP & 1.16 (0.11) & 0.99 (0.1)\\
                         &  \#TP  & 2.39 (0.07) & 2.4 (0.07) & 4.06 (0.15) & 3.95 (0.13)\\
                         & PPV & 0.73 (0.021) & 0.75 (0.022)\\
                         & MSE & 4.49 (0.34) & 3.95 (0.26) & 5.53 (0.3) & 5.09 (0.28)\\
                         \midrule \addlinespace \\
\multirow{3}{*}{50}   & \#FP & 1.05 (0.12) & 0.64 (0.08)\\
                         &  \#TP & 2.92 (0.03) & 2.96 (0.02) & 6.09 (0.12) & 5.89 (0.11)\\
                         & PPV & 0.8 (0.02) & 0.86 (0.016)\\
                         & MSE & 1.32 (0.12) & 1.09 (0.08) & 2.02 (0.10) & 2.04 (0.09)\\
\bottomrule
\end{tabular*}
\caption{Comparison of AdaLasso and STRD-AdaLasso in Examples \ref{table:lasso1} and \ref{table:lasso2}}
\label{table:results_low_ada1}
\end{table}

\begin{table}[!htp]
\footnotesize
\begin{tabular*}{\linewidth}{ @{\extracolsep{\fill}} ll *{16}c @{}}
\toprule
$n$ & Metric & \multicolumn{2}{c}{Example \ref{table:extreme_correlation}} &
\multicolumn{2}{c}{Example \ref{table:independence}} & \multicolumn{2}{c}{Example \ref{table:block}} \\
\midrule \midrule \addlinespace \\
 & & AdaLasso  & STRD-AdaLasso  & AdaLasso  & STRD-AdaLasso  & AdaLasso  & STRD-AdaLasso \\
\addlinespace \\
\multirow{3}{*}{50} & \#FP &2.68 (0.33) & 2.54 (0.2) & 10.36 (0.61) & 8.36 (0.4) & 15.86 (0.52) & 12.64 (0.39)\\
                         &  \#TP  &2.51 (0.1) & 4.27 (0.18) & 6.44 (0.28) & 7.09 (0.21) & 5.8 (0.22) & 6.34 (0.19)\\
                         & PPV &0.65 (0.03) & 0.67 (0.021) & 0.42 (0.017) & 0.48 (0.016) & 0.27 (0.009) & 0.34 (0.012)\\
                         & MSE & 7.22 (0.23) & 6.17 (0.19) & 69.13 (3.80) & 60.34 (3.0) & 60.47 (2.44) & 54.21 (2.26)\\
                         \midrule \addlinespace \\
\multirow{3}{*}{100}   & \#FP &5.29 (0.48) & 2.75 (0.24) & 16.42 (0.67) & 10.12 (0.36) & 18.09 (0.62) & 13.41 (0.5)\\
                         &  \#TP & 5.16 (0.28) & 8.47 (0.13) & 10 (0) & 10 (0) & 9.97 (0.02) & 9.96 (0.02)\\
                         & PPV & 0.62 (0.025) & 0.79 (0.014) & 0.41 (0.013) & 0.52 (0.011) & 0.38 (0.01) & 0.45 (0.01)\\
                         & MSE & 4.81 (0.15) & 2.47 (0.13) & 4.87(0.25) & 3.72 (0.12) & 5.90 (0.31) & 5.85 (0.39)\\
\bottomrule
\end{tabular*}
\caption{Comparison of AdaLasso and STRD-AdaLasso in Examples \ref{table:extreme_correlation}, \ref{table:independence} and \ref{table:block}}
\label{table:results_low_ada2}
\end{table}

\paragraph{Selection}
STRD-Lasso has better false positive control than Lasso/ENet in all cases, and always has similar or better true positive rate than Lasso.  When $p>n$ (Tables \ref{table:independence}, \ref{table:block}), it also achieves similar true positive rate as ENet. In contrast, RLasso in general has good true positive rates, but with poor false positive control. It selects more false positives than STRD-Lasso in all cases. Example \ref{table:extreme_correlation} is the motivating setting for Random Lasso where coefficients of highly correlated variables have different signs. Thanks to group selection property, both STRD-Lasso and RLasso succeed in detecting significantly more correlated relevant variables than other methods, but STRD-Lasso has much better false positive control and higher power than RLasso. 

When $p>n$ and variables are orthogonal to each other (Table \ref{table:independence}), the advantage of false positive control of STRD-Lasso over Lasso, ENet and RLasso is more obvious than in the block correlation design (Table \ref{table:block}). One possible reason is that in Example \ref{table:block}, only one out of 10 correlated variables is relevant in each block, but STRD-Lasso tends to simultaneously select multiple correlated variables, which introduces false positives. 

Comparative performance of AdaLasso and STRD-AdaLasso is broadly similar to that of Lasso and STRD-Lasso. AdaLasso typically has good false positive control, and such advantage is obvious when $p>n$. STRD-AdaLasso typically improves AdaLasso on power and false positive control,  and this is especially true in Example \ref{table:extreme_correlation} where relevant variables with different signs are highly correlated.

\paragraph{Prediction}
STRANDS achieves similar or better MSEs than its base learner in all cases. For the smaller sample sizes ($n=20$ for Examples \ref{table:lasso1}, \ref{table:lasso2} and $n=50$ for Examples \ref{table:extreme_correlation}$\sim$\ref{table:block}), STRD-Lasso's prediction accuracy is notably better than other methods. In ``easier'' settings where sample size is larger ($n=50$ for Examples \ref{table:lasso1}, \ref{table:lasso2} and $n=100$ for Examples \ref{table:extreme_correlation}$\sim$\ref{table:block}), its advantage in prediction is typically less obvious, and can be outperformed by RLasso and AdaLasso when $p>n$ (Examples \ref{table:independence} and \ref{table:block}). 

Example \ref{table:lasso2} is the motivating setup for Elastic Net where the underlying model is not sparse, yet STRD-Lasso manages to achieve comparable prediction error to Elastic Net. Similar to RLasso, STRD-Lasso is favourable when relevant variables with different signs are highly correlated (Example \ref{table:extreme_correlation}). Here, its prediction error is significantly better than Lasso, ENet, and also RLasso (for example, when $n=100$, MSEs of STRD-Lasso and RLasso are 2.21 and 3.18 respectively). The predictive performance of RLasso is more variable, and seems to be competitive when $p>n$ and $n$ is large (see $n=100$ for Examples \ref{table:independence} and \ref{table:block}). 

Comparative performance of AdaLasso and STRD-AdaLasso is broadly similar to that of Lasso and STRD-Lasso. AdaLasso has inferior performance compared to Lasso and ENet when strong multicollinearity exists (see e.g. Example \ref{table:extreme_correlation}), and is competitive for larger $n$. STRD-AdaLasso achieves similar or better predictive performance compared to AdaLasso in all cases, and the improvement is the most significant in Example \ref{table:extreme_correlation} (when $n=100$, the MSEs of AdaLasso and STRD-AdaLasso are 4.81 and 2.47 respectively). 

\paragraph{Null model}
We further compare method performance in a null model, to assess false positive control and bias control. Specifically, we generate data where 300 variables and a response are i.i.d. drawn from $N(\bf{0}, \bf{I})$, so there is no linear relationship between response and any variables. The results are summarised in Table \ref{table:null_model}.

\begin{table}[!h]
\footnotesize
\centering
 \begin{adjustwidth}{-0cm}{}
\begin{tabular}{|c|c|c|c|c|c|c|c|}\hline
$n$ & & Lasso & ENet & AdaLasso & RLasso & STRD-Lasso & STRD-AdaLasso\\\cline{1-8}
 \multirow{2}{*}{$50$}
    & \#FP & 4.53 (0.8) & 6.3 (1.07) & \textbf{2.99 (0.51)} & 20.92 (0.4) & 5.91 (0.5) & 4.41 (0.31)\\\cline{2-8}
    & MSE  & 0.034 (0.007) & \textbf{0.031 (0.006)} & 0.131 (0.018) & 0.285 (0.01) & 0.216 (0.013) & 0.253 (0.013)\\\hline
\multirow{2}{*}{$100$}         
    & \#FP  & 3.07 (0.63)  & 3.41 (0.68) & \textbf{2.08 (0.48)}  & 38.76 (0.61)  & 4.73 (0.47)  & 3.49 (0.39)  \\\cline{2-8}
    & MSE   & 0.011 (0.002)  &  \textbf{0.009} (0.002) & 0.059 (0.012)  & 0.28 (0.01)  & 0.11 (0.01)  & 0.13 (0.011) \\\hline 
\end{tabular}
\end{adjustwidth}
\caption{$p=300$, all variables and response are i.i.d. drawn from $N(\bf{0,I})$}
\label{table:null_model}
\end{table}

Since the screening in STRANDS introduces bias, its has poorer false positive control than its base learner, and prediction error is also inflated, but less severe than RLasso. Since the thresholding rule $1/n$ is somewhat arbitrary, the false positive control of RLasso is rather poor in this case. AdaLasso has the best false positive control, but its prediction is less competitive than Lasso and ENet, since the AdaLasso estimate is biased further away from its initial weights, the Lasso estimate. 

\subsection{High-dimensional settings} \label{subsectio:high_dimensional}
In this section we apply STRANDS and its competitors to some high-dimensional scenarios, to evaluate the performance of STRANDS in high-dimensional settings. We consider the following scenarios with different designs and features. Data are generated using Model \ref{eq:linear_model}. Each variable follows a standard normal distribution, but the correlation structure may vary. All non-zero coefficients have value 3.\\
\\
Example 7: There are $n=200$ samples and $p=500$ variables. All variables are i.i.d. drawn from $\text{N}(\bf{0,I})$. 20 relevant variables are randomly allocated among $p$ variables, SNR=1\\
\\
Example 8: There are $n=100$ samples and $p=500$ variables. The $p$ variables are partitioned into 50 blocks of size 10. Correlation between any pair of variables within the same block is 0.9. Variables in different blocks are independent of each other. There are five relevant variables in each of the first four blocks, and the remaining blocks contain no relevant variables. SNR=4.\\
\\
Example 9: There are $n=300$ samples and $p=1000$ variables. The $p$ variables are partitioned into 10 blocks of size 100. Correlation between any pair of variables within the same block is 0.7. Variables in different blocks are independent of each other. There are two relevant variables in each of the first 10 blocks, and the remaining blocks contain no relevant variables. SNR=2. \\
\\
Example 10: As Example 9, but with Toeplitz correlation design within each block. Specifically, variables $\mathbf{x}_{j_1}$ and $\mathbf{x}_{j_2}$ within the same block have correlation $0.95^{|j_1-j_2|}$. For each pair of two relevant variables in the first 10 blocks, their positions, $j_1'$ and $j_2'$, within a block is chosen such that $|j_1'-j_2'|=7$, to give a correlation of $0.95^7\approx 0.7$.\\

In Example 7 the challenge is high noise level. In Example 8 strong multicollinearity exists. In Examples 9 and 10 a large number of variables are correlated with each other, and the challenge is to find the relevant variables among many correlated irrelevant ones. The only difference is in the correlation structure (pairwise versus Toeplitz). Parameters are tuned or set as before, and we report $\#$TP, $\#$FP, PPV and MSE of each method. The results are summarised in Tables \ref{table:strands_high_dimension_independence} -- \ref{table:strands_high_dimension_large_block_toeplitz}.

\paragraph{Selection}
Across high-dimensional settings, RLasso's true positive rate is among the best, but it has the worst false positive control. This is mainly because the thresholding rule $1/n$ can be too lenient when $n$ is large. In Examples 7 and 8, STRD-Lasso has better false positive control than Lasso and ENet, with similar or better true positive rates. In Example 8 where multiple correlated variables are relevant, ENet has significant gain in TPR over Lasso, and STRD-Lasso also manages to improve upon the true positive rate of Lasso. In Example 9 and 10, Lasso and ENet select many false positives, and STRD-Lasso has similar or slightly worse performance than Lasso with respect to true positive and false positive. The tendency of simultaneously selecting correlated variables may be the reason why STRD-Lasso selects slightly more variables than Lasso in the pairwise correlation design.  

Comparative performance of AdaLasso and STRD-AdaLasso is broadly similar to that of Lasso and STRD-Lasso. AdaLasso typically has good false positive control but less competitive true positive rates. STRD-AdaLasso has similar true positve rates as AdaLasso, and improves false positive control of AdaLasso except in Example 9. 

Overall, we see that STRANDS has less benefit in high-dimensional settings with large correlated blocks and few signals, compared to low-to-moderate dimensional settings. 

\paragraph{Prediction}
Lasso and ENet have similar predictive performance except in Example 8, where relevant variables are strongly correlated, and ENet outperforms Lasso. As expected, both RLasso and STRD-Lasso improves Lasso's prediction accuracy in Example 8, and STRD-Lasso is marginally better than other methods. In other three examples, STRD-Lasso has similar or slightly worse predictive performance than Lasso, and the comparative performance of STRD-Lasso and RLasso varies. RLasso is noticeably worse than STRD-Lasso in the independence design (Example 7, MSEs for STRD-Lasso and RLasso are 115.36 and 122.79). In moderate pairwise correlation design (Example 9), they have similar performance, while in Toeplitz design RLasso achieves the best prediction accuracy among all methods (Example 10).

AdaLasso's predictive performance is among the worst in all settings. This is probably because the initial weights obtained by Lasso are not accurate due to high noise level or strong multicollinearity. STRD-Adalasso has similar or better performance than AdaLasso, and the improvements are significant in data with high noise level (Example 7, MSEs for AdaLasso and STRD-AdaLasso are 149.57 and 128.91) or strong multicollinearity (Example 8, MSEs for AdaLasso and STRD-AdaLasso are 32.38 and 27.34).

Similar to selection performance, we see that the benefit of STRANDS in high-dimensional settings is not as clear as in low-to-moderate dimensional settings. 

\begin{table}[!htp]
\footnotesize
\centering
 \begin{adjustwidth}{-0cm}{}
\begin{tabular}{|r|l|l|l|l|l|l|}
  \hline
 & Lasso & ENet & AdaLasso & RLasso & STRD-Lasso & STRD-AdaLasso \\ 
  \hline
\#FP & 37.36 (2.32) & 44.12 (2.62) & 31.31 (2.04) & 120.23 (1.72) & 30.27 (1.33) & \textbf{25.42 (1.07)} \\ 
\hline
  \#TP & 14.69 (0.38) & 15.2 (0.37) & 14.11 (0.49) & \textbf{18.16 (0.14)} & 14.73 (0.31) & 14.36 (0.29) \\ 
\hline
  PPV & 0.32 (0.015) & 0.29 (0.013) & 0.37 (0.022) & 0.13 (0.0019) & 0.34 (0.0095) & \textbf{0.38 (0.011)} \\ 
\hline
 MSE & \textbf{111.24 (2.64)} & 112.43 (2.40) & 149.57 (4.01) & 122.79 (2.85) & 115.36 (2.89) & 128.91 (3.4) \\ 
   \hline
\end{tabular}
\end{adjustwidth}
\caption{Example 7: $n=200$, $p=500$. All variables are i.i.d. drawn from $\text{N}(\bf{0,I})$. 20 relevant variables are randomly allocated among $p$ variables, SNR=1}
\label{table:strands_high_dimension_independence}
\end{table}

\begin{table}[!htp]
\footnotesize
\centering
 \begin{adjustwidth}{-0cm}{}
\begin{tabular}{|r|l|l|l|l|l|l|}
  \hline
 & Lasso & ENet & AdaLasso & RLasso & STRD-Lasso & STRD-AdaLasso \\ 
  \hline
\#FP & 20.12 (1.03) & 24.56 (1.03) & 11.34 (0.65) & 37.19 (1.09) & 15.72 (0.6) & \textbf{6.62 (0.37)} \\ 
\hline
  \#TP & 15.58 (0.22) & 17.75 (0.19) & 12.23 (0.22) & \textbf{18.25 (0.14)} & 16.08 (0.22) & 12.17 (0.23) \\
\hline 
  PPV & 0.46 (0.015) & 0.44 (0.011) & 0.54 (0.016) & 0.34 (0.0076) & 0.52 (0.012) & \textbf{0.66 (0.014)} \\ 
\hline
 MSE & 28.80 (1.02) & 25.30 (0.88) & 32.38 (1.49) & 24.87 (0.84) & \textbf{24.31 (1.04)} & 27.34 (0.94) \\ 
   \hline
\end{tabular}
\end{adjustwidth}

\caption{Example 8: $n=100$, $p=500$. The $p$ variables are partitioned into 50 blocks of size 10. Correlation between any pair of variables within the same block is 0.9. Variables in different blocks are independent of each other. There are five relevant variables in each of the first four blocks, and the remaining blocks contain no relevant variables. SNR=4.}
\label{table:strands_high_dimension_strong_correlation}
\end{table}

\begin{table}[!htp]
\footnotesize
\centering
 \begin{adjustwidth}{-0cm}{}
\begin{tabular}{|r|l|l|l|l|l|l|}
  \hline
 & Lasso & ENet & AdaLasso & RLasso & STRD-Lasso & STRD-AdaLasso \\ 
  \hline
\#FP & 73.58 (1.49) & 84.53 (1.74) & \textbf{36.2 (1.53)} & 129 (3.59) & 79.81 (1.23) & 39.56 (0.96) \\ 
\hline
  \#TP & 18 (0.16) & 18.23 (0.14) & 16.27 (0.23) & \textbf{18.64 (0.1)} & 17.91 (0.15) & 16.3 (0.24) \\ 
\hline
  PPV & 0.2 (0.0034) & 0.18 (0.003) & \textbf{0.33 (0.0097)} & 0.13 (0.0034) & 0.19 (0.003) & 0.3 (0.0071) \\ 
\hline
 MSE & \textbf{33.14} (0.76) & 33.94 (0.73) & 38.02 (1.25) & 33.98 (0.86) & 34.79 (0.74) & 39.35 (1.07) \\ 
   \hline
\end{tabular}
\end{adjustwidth}

\caption{Example 9: $n=300$, $p=1000$. The $p$ variables are partitioned into 10 blocks of size 100. Correlation between any pair of variables within the same block is 0.7. Variables in different blocks are independent of each other. There are two relevant variables in each of the first 10 blocks, and the remaining blocks contain no relevant variables. SNR=2.}
\label{table:strands_high_correlation_large_block_pairwise}
\end{table}

\begin{table}[!htp]
\footnotesize
\centering
 \begin{adjustwidth}{-0cm}{}
\begin{tabular}{|r|l|l|l|l|l|l|}
  \hline
 & Lasso & ENet & AdaLasso & RLasso & STRD-Lasso & STRD-AdaLasso \\ 
  \hline
\#FP & 58.59 (1.16) & 75.44 (1.42) & 33.06 (1.4) & 93.16 (2.48) & 58.73 (1.06) & \textbf{28.81 (0.82)} \\ 
\hline
  \#TP & 15.36 (0.22) & \textbf{16.97 (0.22)} & 12.7 (0.24) & 16.92 (0.23) & 15.3 (0.24) & 12.88 (0.24) \\ 
\hline
  PPV & 0.21 (0.0041) & 0.19 (0.0032) & 0.29 (0.0082) & 0.16 (0.0033) & 0.21 (0.0039) & \textbf{0.32 (0.0089)} \\ 
  \hline
  MSE & 23.54 (0.53) & 23.57 (0.51) & 26.66 (1.07) &\textbf{21.73 (0.55)} & 24.55 (0.73) & 25.75 (0.75) \\ 
   \hline
\end{tabular}
\end{adjustwidth}

\caption{Example 10: As Example 9, but with Toeplitz correlation design within each block}
\label{table:strands_high_dimension_large_block_toeplitz}
\end{table}

\subsection{Further analysis of STRANDS} \label{section:further_analysis_strands}
\subsubsection{Structural subsampling} \label{subsubsection:structural_subsampling}
To illustrate the advantage of clustering based on correlation, we present the results of two alternative implementations of STRANDS: a) Random Clustering (RC), where the number of clusters $K+1$ and cluster sizes $|G_0| \dots |G_K|$ are the same as in STRANDS, but variables are randomly assigned to each cluster, and b) No Clustering (NC), where all variables are treated as one group. We compare these implementations with the original STRANDS approach (with clustering based on correlation) and with Random Lasso, for Examples \ref{table:extreme_correlation} and \ref{table:block}, which have pairwise correlation structures. Results are shown in Tables \ref{table:extreme_correlation_clustering} and \ref{table:block_clustering}. 

\begin{table}[!htp]
\small
\centering
\begin{tabular}{|c|c|c|c|c|c|c|}\hline
$n$ & &  STRD-Lasso & STRD-Lasso (RC) & STRD-Lasso (NC) & RLasso\\\cline{1-6}
 \multirow{4}{*}{$50$}
    & \#FP & \textbf{3.9 (0.27)} & 4.89 (0.32) & 4.32 (0.33) & 12.91 (0.36)\\\cline{2-6}
    & \#TP & \textbf{6.82 (0.19)} & 4.7 (0.17) & 4.89 (0.15) & 6.6 (0.17)\\\cline{2-6}
    & PPV & \textbf{0.66 (0.019)} & 0.53 (0.019) & 0.58 (0.02) & 0.35 (0.008)\\\cline{2-6}
    & MSE & \textbf{5.22 (0.20)} & 6.68 (0.19) & 6.62 (0.20) & 6.16 (0.16)\\\hline
\multirow{4}{*}{$100$}         
    & \#FP &  \textbf{5.34 (0.41)} & 6.34 (0.45) & 7.03 (0.52) & 16.23 (0.33)\\\cline{2-6}
    & \#TP & \textbf{9.61 (0.08)} & 8.77 (0.13) & 8.86 (0.14) & 9.32 (0.08)\\\cline{2-6}
    & PPV  & \textbf{ 0.69 (0.017)} & 0.63 (0.017) & 0.62 (0.017) & 0.37 (0.004)\\\cline{2-6}
    & MSE & \textbf{2.21 (0.10)} & 3.18 (0.13) & 3.29 (0.14) & 3.18 (0.12)\\\hline
\end{tabular}
\caption{Performance of STRANDS with random clustering (RC) and no clustering (NC) for Example \ref{table:extreme_correlation}. Results of RLasso and STRD-Lasso are the same as in Table \ref{table:extreme_correlation}.}
\label{table:extreme_correlation_clustering}
\end{table}

\begin{table}[!htp]
\small
\centering
\begin{tabular}{|c|c|c|c|c|c|c|}\hline
$n$ & &  STRD-Lasso & STRD-Lasso (RC) & STRD-Lasso (NC) & RLasso\\\cline{1-6}
 \multirow{4}{*}{$50$}
    & \#FP & \textbf{ 23.49 (0.55)} & 28.05 (0.59) & 27.5 (0.59) & 26.8 (0.73)\\\cline{2-6}
    & \#TP & 7.32 (0.17) & \textbf{7.39 (0.16)} & 7.35 (0.17) & 6.45 (0.17)\\\cline{2-6}
    & PPV & \textbf{0.24 (0.0063)} & 0.21 (0.006) & 0.22 (0.006) & 0.2 (0.0067)\\\cline{2-6}
    & MSE & \textbf{51.18 (1.72)} & 52.38 (1.74) & 52.60 (1.8) & 64.86 (1.95)\\\hline
\multirow{4}{*}{$100$}         
    & \#FP & \textbf{35.13 (0.61)} & 41.06 (0.74) & 40.1 (0.67) & 43.62 (0.99)\\\cline{2-6}
    & \#TP & 9.98 (0.01) & \textbf{9.99 (0.01)} & \textbf{9.99 (0.01)} & 9.98 (0.01)\\\cline{2-6}
    & PPV  & \textbf{0.23 (0.003)} & 0.2 (0.003) & 0.2 (0.003) & 0.19 (0.004)\\\cline{2-6}
    & MSE & 9.94 (0.42) & 10.57 (0.43) & 10.58 (0.46) & \textbf{5.26 (0.44)}\\\hline
\end{tabular}
\caption{Performance of STRANDS with random clustering (RC) and no clustering (NC) for Example \ref{table:block}. Results of RLasso and STRD-Lasso are the same as in Table \ref{table:block}.}
\label{table:block_clustering}
\end{table}

For data with clear correlation structure, the clustering step of STRANDS can significantly improve its performance. In Example \ref{table:extreme_correlation}, RC and NC both lead to much worse TP, FP, PPV and MSE (Table \ref{table:extreme_correlation_clustering}), and in Example \ref{table:block} they lead to more FP, inferior MSE and similar TP, although the deterioration in performance is not as striking (Table \ref{table:block_clustering}). 

\begin{figure}[!htp]
\centering
\includegraphics[height=4in]{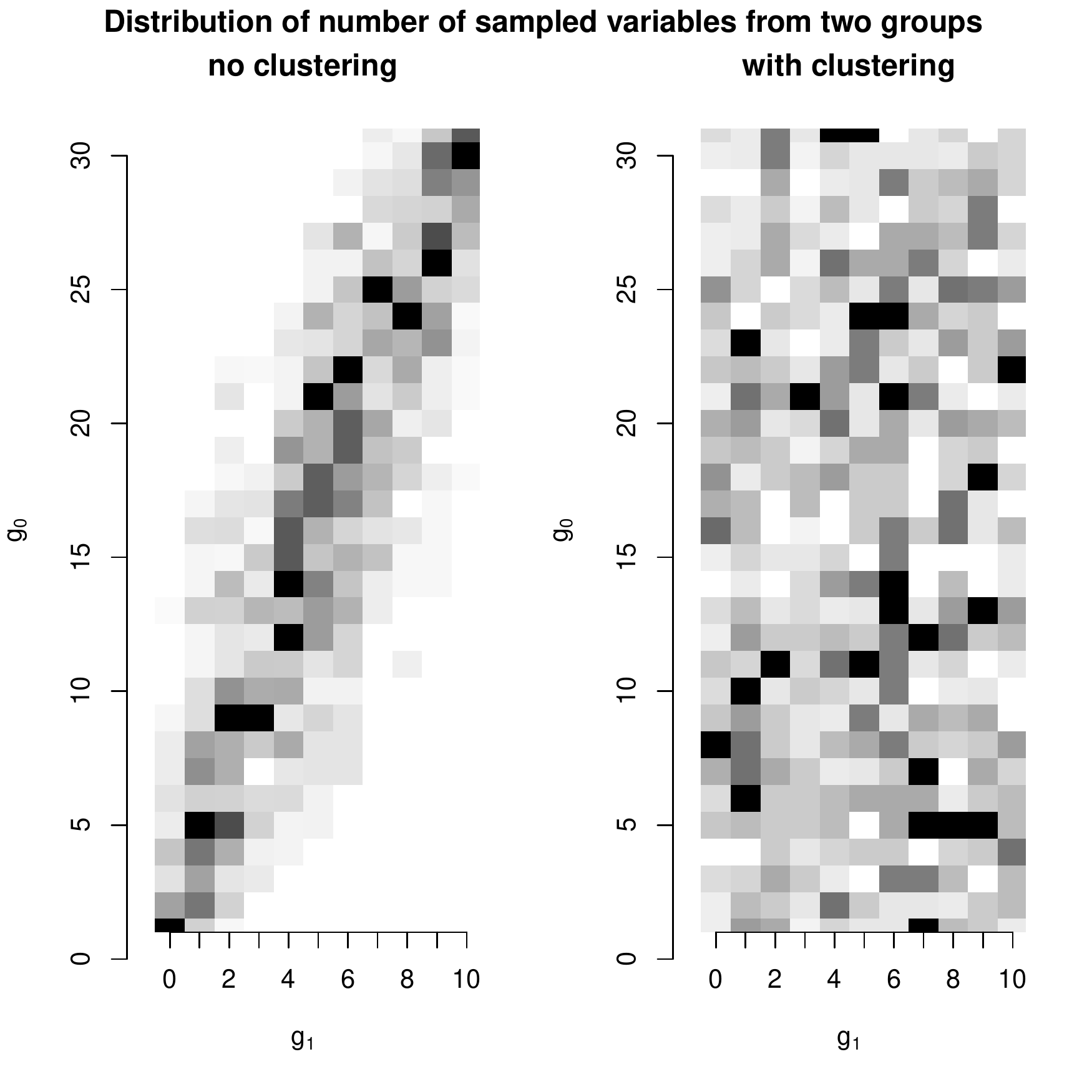}
\caption{Influence of the clustering step (Step 0 of STRANDS), on the sampling of variables (Step 1a (i)) for Example \ref{table:extreme_correlation}, where there are two groups. Let $G_0$ be the group of 30 independent variables and $G_1$ be the group of 10 highly correlated variables. Left panel shows the result without clustering step (i.e., all variables are treated as one group) and right panel shows the result with clustering step. Sampling of variables is repeated 1000 times, and each time corresponds to a combination of $g_0$ (number of variable sampled from $G_0$, y axis) and $g_1$ (number of variables sampled from $G_1$, x axis). Darker colour indicates that the combination of certain $g_0$ and $g_1$ appears more frequently than others.}
\label{fig:benefit_of_clustering}
\end{figure}

The purpose of clustering is to explore more efficiently the model space, rather than only exploring models with certain characteristics. This is illustrated in Figure \ref{fig:benefit_of_clustering}, using the pairwise correlation structure of Example \ref{table:extreme_correlation}. The data has one correlated group of size 10, and one independent group of size 30. Without clustering step, we sample from the set of all variables, and the ratio of numbers of sampled variables from the two groups will be close to the ratio of the corresponding group sizes, i.e., 1:3 in this case. So we are likely to focus on a small regime of model space. With clustering, we sample at random from each group independently, and all combinations of variables from different groups are equally likely, so the exploration in Step 1 of STRANDS is more comprehensive and effective. 

\subsection{Application to real data}
\label{sec:real_data}
In this section we analyse three real data examples by using STRANDS and other variants of Lasso. All variables are standardised to have mean zero and variance one. We first run all methods on the full dataset to get model sizes, and then compare their predictive performance as follows: random split the dataset into a training dataset ($\boldsymbol{X_{train}}$, $\boldsymbol{Y_{train}}$) with $90\%$ of samples and a test dataset ($\boldsymbol{X_{test}}$, $\boldsymbol{Y_{test}}$) of the other $10\%$ of samples ; models are fitted on the training data and the coefficient estimates $\hat{\boldsymbol{\beta}}$ are used to predict the response of the test data, \begin{equation}
\text{Prediction Error}=||\boldsymbol{Y_{test}}-\boldsymbol{X_{test}}\hat{\boldsymbol{\beta}}||_2^2/n_{test}.
\end{equation} We applied Lasso, Elastic Net (ENet, $\alpha=0.5$), Adaptive Lasso (AdaLasso), Random Lasso (RLasso), STRANDS-Lasso (STRD-Lasso), and STRANDS-Adaptive Lasso (STRD-AdaLasso) to fit linear regression models. Parameters and tuning strategies are the same as in simulation study. Data splitting is repeated 100 times, and mean prediction errors across the 100 times are reported, with bootstrap standard errors in parentheses. We also report the computation time of every method for running on the full data. In all examples STRD-Lasso takes significantly less computation time than RLasso (STRD-Lasso is 3-4 times faster than RLasso). 
 
\subsubsection{Gene expression and aging}
The first dataset is from \citet{Lu2004}. The study aims to discover the relationship between aging and gene expression levels in human frontal cortex. The response is ages of $n=30$ patients (ranging from 26 to 106 years), and explanatory variables are expression levels of 12,625 genes measured by microarray. Linear model is built to find those genes that can predict the age of a patient. Gene expression values are log-transformed, and they filter out genes based on local false non-discovery rates. After preprocessing, $p=403$ variables are retained as candidate variables. This dataset has previously been analysed in \cite{Zuber2011}, and is available in R package \textit{care}. As in \cite{Zuber2011}, the response is standardised.

The results of aging data analysis are summarised in Table \ref{table:real_data1}. We see that STRANDS-Lasso achieves the smallest mean prediction error, followed by Elastic Net and Lasso. Prediction errors of Adaptive Lasso and Random Lasso are less satisfactory, perhaps because they both have overly sparse models and missed more signals. For Random Lasso, this could be due to the fact that sample size of training data $n=27$ is small, so that the coefficient estimate threshold $1/n$ is too restrictive and the resulting model is too conservative. On the other hand, STRANDS-Lasso and STRANDS-Adaptive Lasso have similar model sizes as their base learners, with improved prediction errors respectively.

\begin{table}[!htp]
\small
\centering
 \begin{adjustwidth}{-0cm}{}
\begin{tabular}{|r|l|l|l|l|l|l|}
  \hline
 & Lasso & ENet & AdaLasso  & RLasso & STRD-Lasso & STRD-AdaLasso\\ 
  \hline
Model Size  & 23 & 29 & 9 & 13 & 21 & 10 \\ 
\hline
\begin{tabular}{@{}c@{}}Prediction \\ Error\end{tabular}   & \begin{tabular}{@{}c@{}}0.344 \\ (0.028)\end{tabular}  & \begin{tabular}{@{}c@{}}0.313 \\(0.024)\end{tabular}  & \begin{tabular}{@{}c@{}}0.465 \\ (0.042)\end{tabular}   & \begin{tabular}{@{}c@{}}0.563 \\ (0.036) \end{tabular}  & \begin{tabular}{@{}c@{}}0.291 \\ (0.019) \end{tabular}  & \begin{tabular}{@{}c@{}}0.429 \\ (0.031)\end{tabular} \\ 
\hline
Time (s)  &<1 & <1 & <1  & 205.33  & 53.24  & 102.15\\
\hline

\end{tabular}
\end{adjustwidth}
\caption {Real data analysis : age data, $n=30, p=403$}
\label{table:real_data1}
\end{table}

\subsubsection{Gene expression and Bardet-Biedl syndrome} 
Next we consider the dataset of mammalian eye tissue, reported in \cite{Scheetz2006}. Tissue harvesting is applied for 120 male rats' eyes, followed by microarray analysis. Expression levels of over 28,000 genes are measured. It is known that TRIM32 gene is the cause of Bardet-Biedl syndrome \citep{Chiang2006}, and linear regression is applied to find genes most related to TRIM32. It is believed that only a small number of genes are associated with TRIM32, and after pre-processing, the resulting dataset consists of 200 variables, which is available in R package \textit{flare}. 

The results of eye data analysis are summarised in Table \ref{table:real_data2}. STRANDS-Lasso again achieves the smallest mean prediction error, followed by Elastic Net and Lasso. Adaptive Lasso is overly sparse and worst for prediction purpose, while Random Lasso does not do well for prediction either, and selects less variables than Lasso. Unlike in age data, STRANDS-Lasso selects more variables than Lasso, similar to Elastic Net. This could be due to the fact that strong multicollinearity exists in eye data, where the mean absolute pairwise correlation among the 200 variables is 0.606. Group selection property of STRANDS leads to larger model, with more accurately estimated coefficients. STRANDS-Adaptive Lasso also significantly improves the prediction performance over Adaptive Lasso, with larger model.  

\begin{table}[!htp]
\small
\centering
 \begin{adjustwidth}{-0cm}{}
\begin{tabular}{|r|l|l|l|l|l|l|}
  \hline
& Lasso & ENet & AdaLasso  & RLasso  & STRD-Lasso & STRD-AdaLasso\\ 
  \hline
Model Size & 26 & 31 & 3  & 18  & 33 & 10\\ 
\hline
\begin{tabular}{@{}c@{}}Prediction \\ Error ($\times 10^{-3}$)\end{tabular}   & \begin{tabular}{@{}c@{}}9.23 \\ (0.63)\end{tabular}  & \begin{tabular}{@{}c@{}}9.01 \\ (0.60) \end{tabular}  & \begin{tabular}{@{}c@{}}14.08 \\ (1.66)\end{tabular}   & \begin{tabular}{@{}c@{}}12.05 \\ (1.07)  \end{tabular}  & \begin{tabular}{@{}c@{}}8.76  \\ (0.48)\end{tabular}  &\begin{tabular}{@{}c@{}}9.59 \\ (0.72)\end{tabular} \\ 
\hline
Time (s)  &<1 & <1 & <1  & 221.55  & 67.31 & 121.34\\
\hline

\end{tabular}
\end{adjustwidth}
\caption {Real data analysis : eye data, $n=120, p=200$}
\label{table:real_data2}
\end{table}

\subsubsection{Plasma proteome analysis}   
Finally we compare STRANDS and its competitors in an analysis of plasma proteome data from \cite{Kirk2011}. Human T lymphotropic virus Type 1 (HTLV-1) is one of the main causes of certain chronic inflammatory diseases, which are commonly referred to as HTLV-1-associated myelopathy/tropic spastic paraparesis (HAM). It is also responsible for adult T cell leukaemia/lymphoma. The study uses surface-enhanced laser desorption mass spectrometry technique to analyse the plasma proteome of 65 HTLV-1-infected patients, who either have HAM disease or are asymptomatic HTLV-1 carriers (AC). The dataset contains log intensities of 49 protein peaks (represented by m/z value in kDa), and the responses are log proviral load and disease status of patients (either HAM or AC). \citet{Kirk2011} and \citet{Kirk2013} used binary disease status as the response and built logistic regression model to find the peaks that most significantly contribute to the discrimination of disease outcomes. We use log proviral load as the response and build linear regression model to find the associations between log proviral load and protein peaks. Since proviral load and disease outcome are closely related, our analysis can be regarded as a continuous relaxation of the binary problem. 

The results of plasma proteome analysis are summarised in Table \ref{table:real_data3}. Prediction errors are similar across all methods. Adaptive Lasso has the worst predictive performance, and Random Lasso and STRANDS-Adaptive Lasso are marginally better than others. Random Lasso has the largest model size, followed by Elastic Net and Lasso. STRANDS again improves over its base learners, and the improvement is marginal for Lasso, but significant for adaptive Lasso. STRANDS also provides sparser results than its base learners in this example. 

For selection, STRANDS-Lasso selects the following peaks: 6.83, 10.8, 11.7, 12.2, 15.7, 17.2, 17.5 and 25.4 kDa; Lasso selects one more peak, 59.5 kDa. We find that peaks selected by linear regression and logistic regression (see below) coincide to some degree. \citet{Kirk2013} apply seven different selection strategies using sparse logistic regression models, and selected protein peaks 10.8, 11.7, 11.9, 13.3, 14.6 17.3, 17.5 and 25.1 kDa. \cite{Kirk2011} applied univariate analysis and logistic regression with Lasso penalty, and selected peaks 11.7, 11.9, and 13.3 kDa. In both studies the authors claimed that 11.7 and 13.3 kDa peaks have the strongest signal, and contribute significantly to predicting patient disease status. Both liner regression and logistic regression select 10.8, 11.7, and 17.5 peaks, and the new peaks selected by linear regression are also interesting and could be investigated further.

\begin{table}[!htp]
\small
\centering
 \begin{adjustwidth}{-0cm}{}
\begin{tabular}{|r|l|l|l|l|l|l|}
  \hline
& Lasso & ENet & AdaLasso  & RLasso  & STRD-Lasso & STRD-AdaLasso\\ 
  \hline
Model Size & 9 & 12 & 6 & 15 & 8 & 5\\ 
\hline
\begin{tabular}{@{}c@{}}Prediction \\ Error\end{tabular}  & \begin{tabular}{@{}c@{}}0.573 \\ (0.028)\end{tabular}  &\begin{tabular}{@{}c@{}}0.568 \\ (0.028)\end{tabular}   & \begin{tabular}{@{}c@{}}0.634 \\ (0.033)\end{tabular}   &  \begin{tabular}{@{}c@{}}0.557 \\ (0.03) \end{tabular}  & \begin{tabular}{@{}c@{}}0.566 \\ (0.032) \end{tabular} & \begin{tabular}{@{}c@{}}0.554 \\ (0.028)\end{tabular} \\ 
\hline
Time (s)  &<1 & <1 & <1  & 264.14 & 69.19 & 101.45\\
\hline

\end{tabular}
\end{adjustwidth}
\caption {Real data analysis : plasma proteome data, $n=65, p=49$}
\label{table:real_data3}
\end{table}

\section{Conclusion and Discussion} \label{section:discussion}
Recent developments in biological science enables researchers to collect massive amount of omics data, which is incredibly informative for learning patterns and dynamics in biological systems. Major obstacles of using such data include high dimensionality, spurious random correlations, high noise level, measurement errors, among others. Robustness becomes a key factor when modelling omics data due to large uncertainty. There may not be definitive evidence for a single optimal model and evidence can be distributed among different candidate models. Ensemble learning is a powerful tool in this situation, where each candidate model serves as a weak learner, and combining them in a strategic way yields a stronger model than each individual one. In this paper we proposed a new variable selection strategy, structural randomised selection (STRANDS), in line with Random Lasso. The method inherits some properties from Random Lasso and alleviates some issues of $l_1$ penalised regression. Particularly, both STRANDS and Random Lasso encourage strongly correlated variables to be selected simultaneously, and the model size is no longer limited by the sample size. STRANDS takes into account the correlation structure to more effectively explore the model space. Compared to Random Lasso, STRANDS's implementation is more automatic such that less parameters need to be tuned, so the method is computationally less intensive and easy to be parallelised. Without bootstrap step, STRANDS uses full sample information in all candidate models. Both simulation study and real data study compared the performance of STRANDS with popular regularised regression methods. We found that STRANDS typically improves the performance of its base learner in low-to-moderate dimensional settings, especially when strong multicollinearity exists; in high-dimensional settings the benefit of STRANDS is less clear. Although in this paper the baseline learners used with STRANDS were Lasso and Adaptive Lasso, STRANDS can also be combined with other sparse regression methods to improve their performance.


Both Random Lasso and STRANDS can be interpreted as randomised screening methods. Screening is a supervised approach to reduce dimensionality, where irrelevant variables are eliminated according to their relationship with the response. As a pioneering work, \cite{Fan2007} proposed to use univariate regression coefficients to reduce the dimensionality of ultra-high dimensional data. Specifically, their Sure Independence Screening (SIS) approach first screens the data to have only a number of variables with largest univariate regression coefficients, and subsequent modelling analysis is then performed on the sub-data. The method is very straightforward and is proved to have sure screening property in an asymptotic framework, i.e., the probability of screened sub-data containing all relevant variables converges to one as the sample size approaches infinity. They also propose an Iterative Sure Independence Screening (ISIS) method to enhance the finite sample performance. However, there are two potential issues with this method. Firstly, the univariate regression coefficient can be highly unstable to represent the actual importance of a variable, especially in high-dimensional data with complicated correlation structure. Secondly, there is no justification on how to choose the degree of screening, i.e., how many variables should be retained in the sub-data. In Random Lasso, instead of univariate regression, the importance measure is calculated by $l_1$ penalisation method on random combination of variables, which captures the multivariate effects and sparsity pattern of data; instead of arbitrary screening degree, Random Lasso tunes the number of screened variables in a data-driven way, although the procedure can be computationally intensive. STRANDS addresses these two issues in an adaptive and data-driven way. As in Random Lasso, we explore variable importance by repeatedly applying penalised regression on random subsets of variables in Step 1, to take into account the multivariate effects instead of univariate effects as in SIS. Multivariate effects are more reliable when dimensionality is high and correlation structure is complex. The degree of screening is automatically obtained based on selection probabilities from Step 1, which is computationally efficient and more accurate.

\section*{Code availability}
The scripts for implementing STRANDS (Lasso as base learner) is available at \url{https://github.com/fw307/structural_randomised_selection}. 

\begin{subappendices}
\section*{Supplementary results} \label{sec:appendix}
\renewcommand{\thefigure}{.S\arabic{figure}}
\end{subappendices}

\subsection*{Two-step variable sampling}
\begin{figure}[!htp]
\centering
\includegraphics[height=4in]{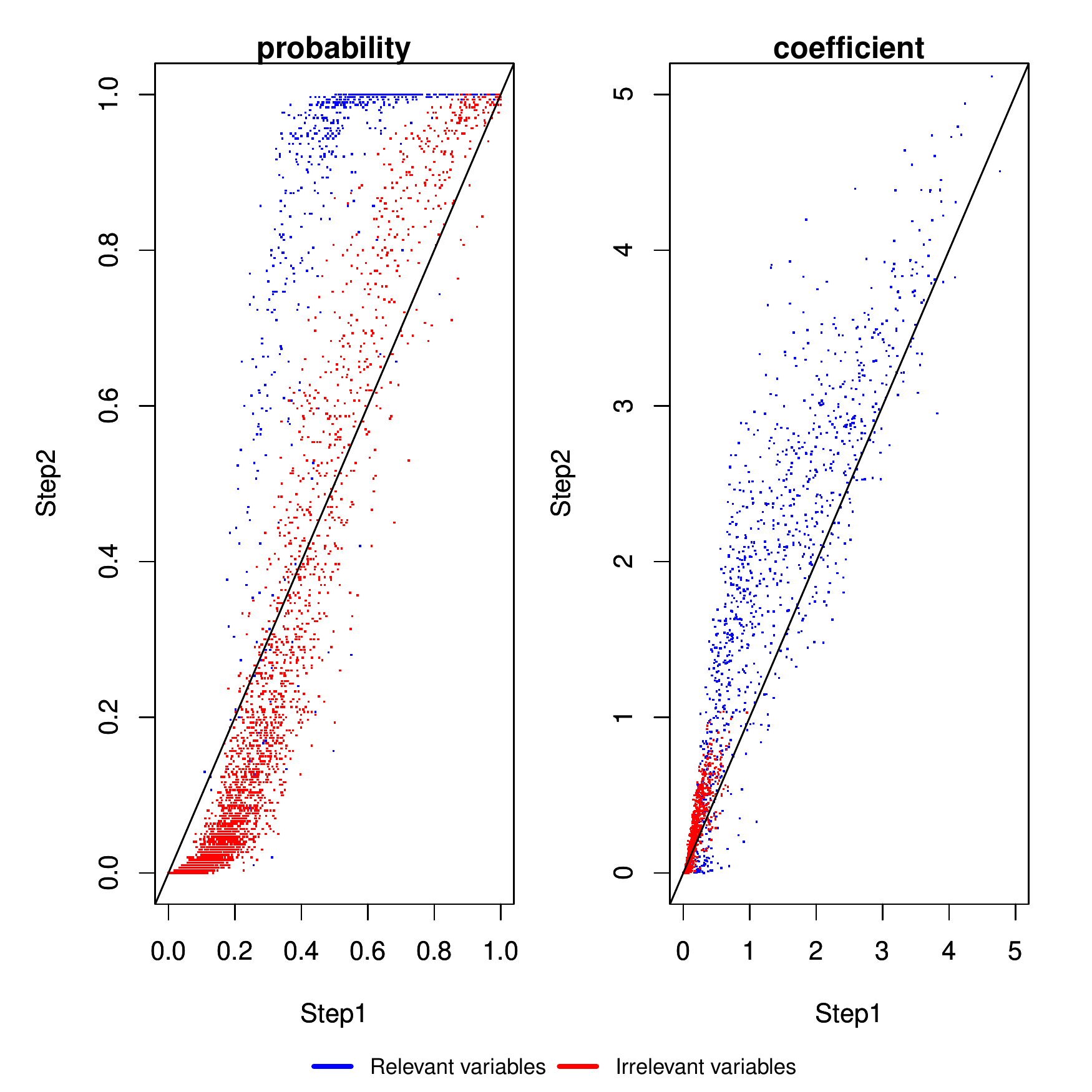}
\caption{Comparison of importance scores from Step 1 and 2 of STRANDS-Lasso for Example \ref{table:extreme_correlation}, as in Table \ref{table:extreme_correlation}, $n=100$. Simulation is repeated 100 times, and each dot corresponds to a pair of scores (probability or absolute coefficient in Step 1 and Step 2) of a variable in one simulation. Left panel shows the comparison of selection probabilities from Step 1 ($\theta_j$) and Step 2 ($\hat{\pi}_j$), and right panel shows the comparison of (absolute) coefficient estimates from Step 1 ($\alpha_j$) and Step 2 ($|\hat{\beta}_j|$). Relevant variables and irrelevant variables are presented by blue and red dots respectively.}
\label{fig:strands_step12}
\end{figure}

In Step 1 STRANDS samples in a uniform way from each group of variables obtained in the clustering step, and applies the base learner algorithm to the random subsets of variables. Although the averaged results on these random subsets of variables do not provide accurate coefficient estimates since candidate models can be mis-specified, they offer guidance on relative importance of variables and underlying sparsity of the true model. Step 2 exploits the relative importance by random sampling, such that variables with larger importance scores from Step 1 are more likely to be sampled. Thus relevant variables are more likely to be included in candidate models while irrelevant one are more likely to be excluded. Step 2 evaluates joint effect of relevant variables by applying the base learner algorithm, and calculates importance scores, which are more accurate than those in Step 1. This is illustrated in Figure \ref{fig:strands_step12}. We use Example \ref{table:extreme_correlation} with clear correlation structure, and use Lasso as the base learner for STRANDS, to show how the importance scores change between Step 1 and Step 2. Left panel shows the comparison of selection probabilities of Step 1 and Step 2, and right panel shows the comparison of (absolute) coefficient estimates. We see that selection probabilities of relevant variables are boosted after importance-score-guided sampling in Step 2, and selection probabilities of irrelevant variables are typically down-weighted. Similarly, coefficient estimates of relevant variables are also improved after Step 2. This justifies the necessity of Step 2 in STRANDS.  \\

\newpage
\bibliographystyle{apalike}
\renewcommand{\bibname}{References} 
\bibliography{Record} 

\end{document}